\documentclass[%
 reprint,
 superscriptaddress,
 amsmath,amssymb,
 aps,
floatfix,
prx]{revtex4-1}

\usepackage[utf8]{inputenc}  
\usepackage{graphicx}
\usepackage{tikz}
\usetikzlibrary{matrix}
\usetikzlibrary{positioning,shapes}
\usepackage{bm}
\usepackage{MjEpp}
\usepackage{cancel}

\newcommand{\QSERG}{Quantum Systems Engineering Research Group, Department of Physics, Loughborough University, Loughborough, LE11 3TU, United Kingdom}
\newcommand{\Wolfson}{The Wolfson School, Loughborough University, Loughborough, LE11 3TU, United Kingdom}

\newcommand{\SpatialParity}{\hat{\Pi}}
\newcommand{\SpatialKernel}{\hat{\Pi}}
\newcommand{\SpinParity}{\hat{\pi}}
\newcommand{\SpinKernel}{\hat{\pi}}
\newcommand{\ElectronParity}{\hat{\Pi}^{\mathrm{e^-}}}

\newcommand{\SpatialArgs}{\mathbf{q}_i,\mathbf{p}_i}

\newcommand{\lettering}[1]{\LARGE\textsf{\textbf{\textcolor{black}{#1}}}}
\newcommand{\letteringW}[1]{\LARGE\textsf{\textbf{\textcolor{white}{#1}}}}
\newcommand{\letterining}[1]{\large\textsf{\textbf{\textcolor{black}{#1}}}}

\begin{document}
\title{Visualizing entanglement in atoms and molecules}
\author{B.~I.~Davies}
\affiliation{\QSERG}
\author{R.~P.~Rundle}
\affiliation{\QSERG}
\affiliation{\Wolfson}
\author{V.~M.~Dwyer}
\affiliation{\QSERG}
\affiliation{\Wolfson}
\author{J.~H.~Samson}
\affiliation{\QSERG}
\author{Todd Tilma}
\affiliation{Tokyo Institute of Technology, 2-12-1 Oookayama, Meguro-ku, Tokyo 152-8550, Japan}
\affiliation{\QSERG}
\author{M.~J.~Everitt}
\email{m.j.everitt@physics.org}
\affiliation{\QSERG}
\date{\today}
%
\begin{abstract}
In this work we show how constructing Wigner functions of heterogeneous quantum systems leads to new capability in the visualization of quantum states of atoms and molecules. 
This method allows us to display quantum correlations (entanglement) between spin and spatial degrees of freedom (spin-orbit coupling) and between spin degrees of freedom, as well as more complex combinations of spin and spatial entanglement for the first time. 
This is important as there is growing recognition that such properties affect the physical characteristics, and chemistry, of atoms and molecules.
Our visualizations are sufficiently accessible that, with some preparation, those with a non-technical background can gain an appreciation of subtle quantum properties of atomic and other systems.
By providing new insights and modelling capability, our phase-space representation will be of great utility in understanding aspects of atomic physics and chemistry not available with current techniques.
\end{abstract}
\maketitle
%
\section{Introduction} 
Despite its fundamental difficulties, the Rutherford description of the atom as electrons orbiting a nucleus is an established icon of the physical sciences. 
This provides a familiar image with which to start a discussion of matter at the subatomic level. 
In such discussions one rapidly moves towards a more sophisticated view of a set of atomic and molecular orbitals, generally displayed as the 90-percentile of the probability density of the associated quantum-mechanical energy eigenstate. 
These images represent a much more accurate view; however, some simplifications remain.
For example, they are unable to display the entanglement of spin and spatial degrees of freedom due to coupling between the spin of an electron and its orbital angular momentum. 
This spin-orbit coupling contains key features that change the shape of an energy eigenstate as well as affecting chemical properties such as dissociation energy~\cite{Roos2004,Boguslawski2015, Dehesa2012,Esquivel2011}. 
Given the growing recognition that phenomena such as spin-orbit coupling play an important role in some chemical reactions~\cite{Miller2019,Zobel2018,Li2019}, there is a need for tools to help better understand these processes.

In this work we bring insight to atomic systems by presenting a framework for visualizing states such as those found using modern quantum-chemistry numerical simulations (which include both spin and entanglement~\cite{Amovilli2004,Coe2008,Osenda2007,Pipek2009}).
To do this we extend the standard picture of the probability density to the full atomic phase space, including spin degrees of freedom.
Whilst there have been a number of previous attempts to visualize atoms using these techniques, none have so far included spin~\cite{Besley2003,Besley2004,Besley2005,Gill2003,Gill2006}.
Representing atoms and molecules in phase space (via Wigner functions) allows for a complete description of the quantum state as a quasi-probability density function.
While~\Refs{PhysRevLett.117.180401,Rundle2019} lay down the necessary framework for heterogeneous systems (by which we mean systems combining differing continuous phase space representations).
we are aware of only two other examples considering the Wigner functions of heterogeneous quantum systems completely within phase space.
One considers using the Wigner function as an entanglement witness for hybrid bipartite states~\cite{Arkhipov2018}. 
The other~\cite{RundleArx2018} investigates the phase-space representation of one or more two level systems coupled to a cavity mode in the Jaynes- and Tavis-Cummings models.
Our simple procedure however, allows for the construction of Wigner functions of composite heterogeneous systems.

We demonstrate below how such methods can be used to visualize spin-orbital, spin-spin, as well as other more complex entanglement combinations of spin and spatial degrees of freedom. 
We expect that this capability will find great utility in understanding important electronic transfer processes; such as photosynthesis (PSI and PSII), the avian compasses and oxygen transport via hemoglobin in blood~\cite{Bradler2010, Cai2010,Cai2010a, Caruso2010,Novoselov2016,Sarovar2010}. 
Having said this, spin-orbital entanglement is not trivial, particularly for many-electron systems.
It is with these future applications in mind that we demonstrate a more accurate visualization of the atom; one that is familiar, yet at the same time offers more insight into the internal entanglement effects that determine many atomic properties~\cite{Boguslawski2015,Dehesa2012, Esquivel2011,Osenda2007, Tichy2011}.
%
\section{Particles in phase space}
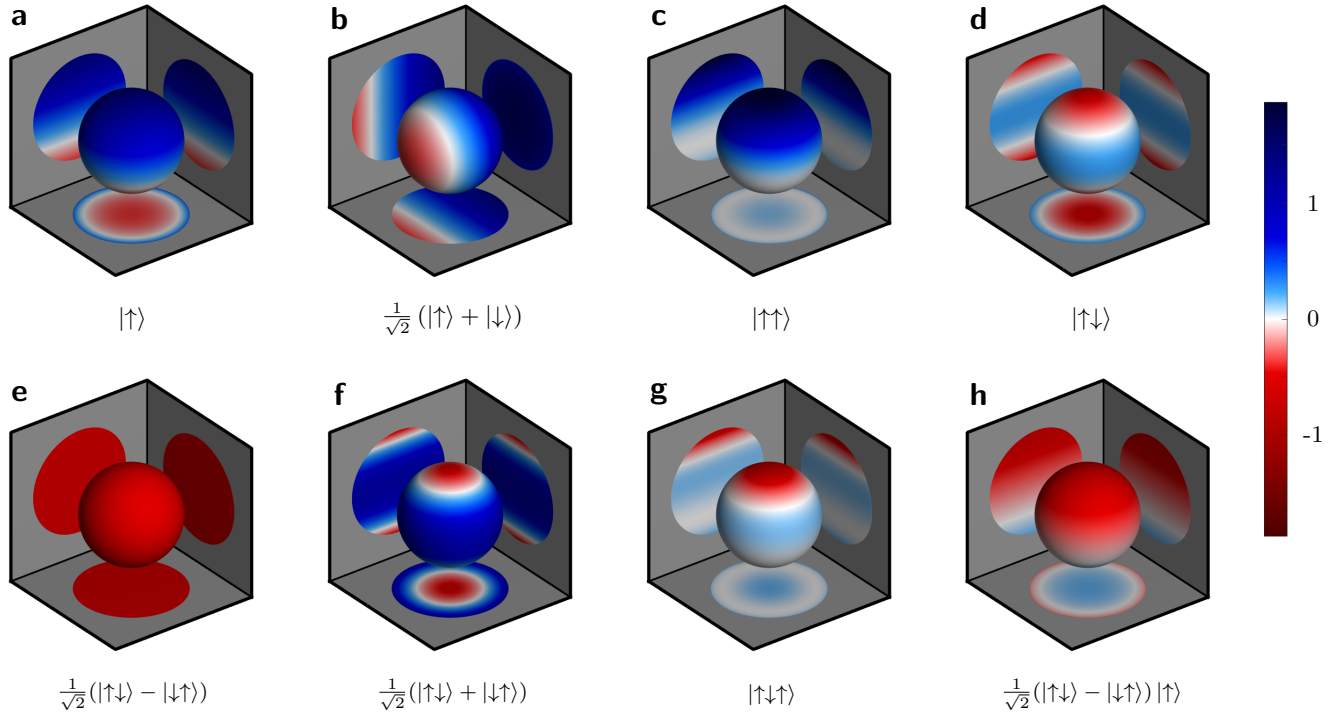
\begin{figure*}[!tb]
	\begin{tikzpicture}		
        \node at (0,0) {
        \includegraphics[width = \linewidth, trim = {13cm, 5cm, 0cm, 2cm}, clip = true]{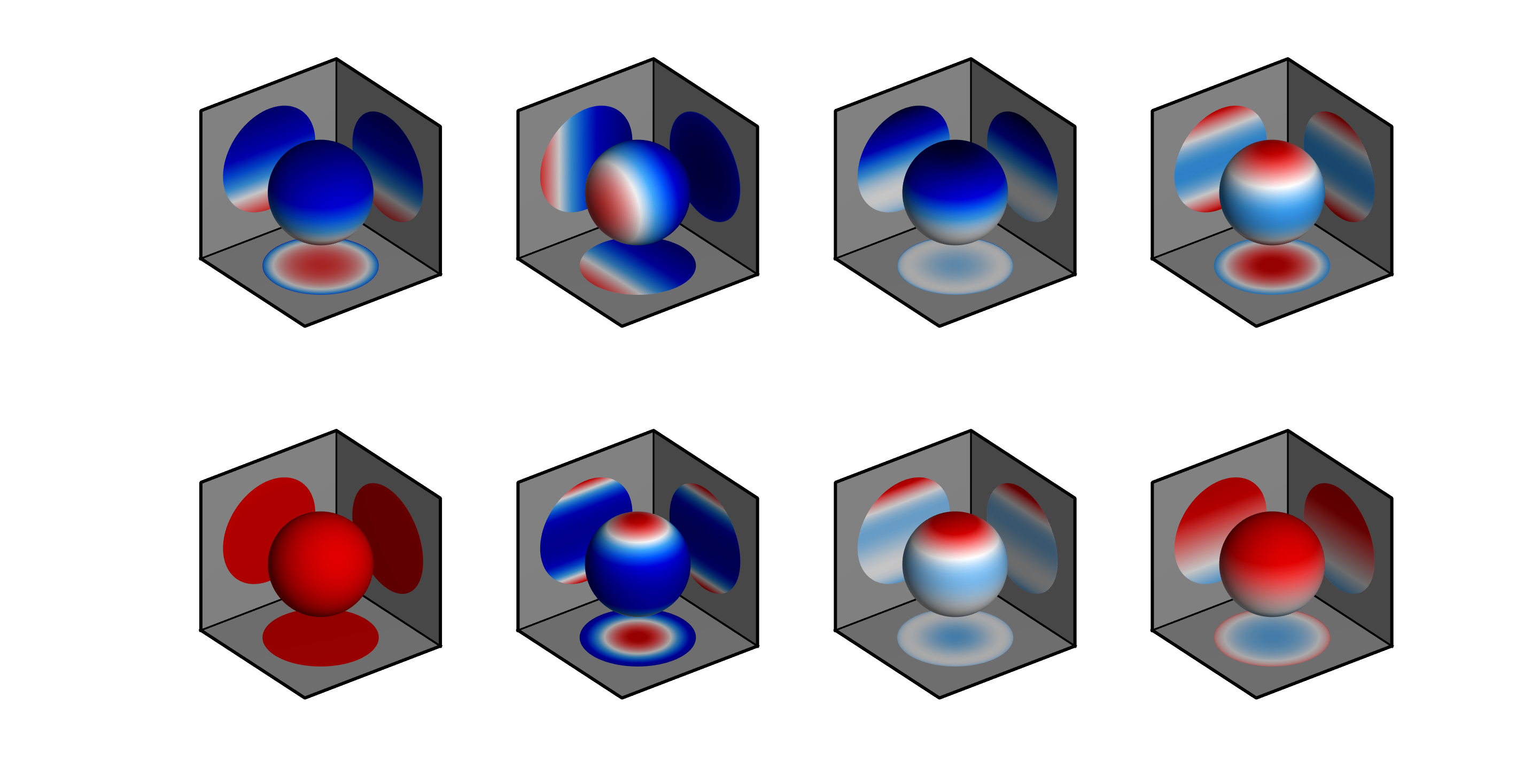}
        };        
        
        \def\cbx{7.9}
        \node at (\cbx,0) {
        	\includegraphics[width = 0.3cm]{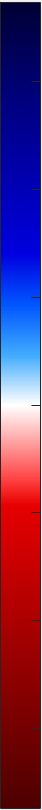}
        };
        \node at (\cbx+0.5,0) {0};
        \node at (\cbx+0.5,1.54) {1};
        \node at (\cbx+0.5,-1.54) {-1};
        \def\topLettY{ 4.0}
        \def\botLettY{-1.0}
        \def\lettX{-8.8}
        \def\lettDispX{4.25}
        \node at (\lettX, \topLettY) 
            {\letterining{a}};
        \node at (\lettX+\lettDispX, \topLettY) 
            {\letterining{b}};
        \node at (\lettX+2*\lettDispX, \topLettY) 
            {\letterining{c}};
        \node at (\lettX+3*\lettDispX, \topLettY) 
            {\letterining{d}};
        \node at (\lettX, \botLettY) 
            {\letterining{e}};
        \node at (\lettX+\lettDispX, \botLettY) 
            {\letterining{f}};
        \node at (\lettX+2*\lettDispX, \botLettY) 
            {\letterining{g}};
        \node at (\lettX+3*\lettDispX, \botLettY) 
            {\letterining{h}};
        \def\txtX{-7.3}
        \node[anchor = center] at (\txtX,0.0) 
        {$\Up$};
        
        \node[anchor = center] at (\txtX+\lettDispX,0.0) 
        {$\frac{1}{\sqrt{2}}\left(\Up + \Down\right)$};
        
        \node at (\txtX+2*\lettDispX,0.0) 
        {$\ket{\uparrow\uparrow}$};
        
        \node at (\txtX+3*\lettDispX,0.0) 
        {$\ket{\uparrow\downarrow}$};
        
        \node at (\txtX,-5) 
        {\footnotesize $\frac{1}{\sqrt{2}}(\ket{\uparrow\downarrow}-\ket{\downarrow\uparrow})$};
        \node at (\txtX+\lettDispX,-5) 
        {\footnotesize $\frac{1}{\sqrt{2}}(\ket{\uparrow\downarrow}+\ket{\downarrow\uparrow})$};
        \node at (\txtX+2*\lettDispX,-5) {\footnotesize $\ket{\uparrow\downarrow\uparrow}$};
        \node at (\txtX+3*\lettDispX,-5) 
        {\footnotesize $\frac{1}{\sqrt{2}}(\ket{\uparrow\downarrow}-\ket{\downarrow\uparrow})\ket{\uparrow}$};
	\end{tikzpicture}
\caption{
A set of reference plots of spin Wigner functions to aid interpretation of the results presented later in this work. The state vectors for each Wigner function are given under each image.
Multi-spin states have been plotted on the equal angle slice, $\theta_i=\theta$ and $\phi_i=\phi$ for all $i$. 
Note that~\Sub{c} is the product of two states which individually are the same as~\Sub{a}, \Sub{g} is the product of~\Sub{a} and~\Sub{d}, and~\Sub{h} is the product of~\Sub{a} and~\Sub{e}.
See~\Ref{Rundle2016} for a full discussion. 
} \label{Fig:Ref}
\end{figure*}
It is possible to write the state of any system as a quasi-probability distribution over the system's degrees of freedom~\cite{PhysRevLett.117.180401,Rundle2016,Rundle2019}.
This is termed the Wigner function and can be calculated by taking the expectation value of a suitably displaced parity operator over all its possible configurations (the phase space). 
For the electron this generalized parity is the tensor product of the displaced spatial parity $\SpatialKernel_i(\SpatialArgs)$ and a generalized displaced spin  parity $\SpinKernel_i(\theta_i,\phi_i)$.
\be
    \ElectronParity_i(\SpatialArgs,\theta_i,\phi_i)
        =
    \SpatialParity_i(\SpatialArgs) \otimes \SpinParity_i(\theta_i,\phi_i).
\ee
The spatial parity $\SpatialParity$ is the operator that reflects states through the origin in phase space, displaced by the displacement operator $\hat D_i(\SpatialArgs) = \exp\left(\ui[\mathbf{p}_i\cdot \hat{\mathbf{q}}_i - \mathbf{q}_i \cdot \hat{\mathbf{p}}_i \right]/\hbar)$ so that $\SpatialKernel_i(\SpatialArgs)= \hat D_i(\SpatialArgs) \SpatialParity \hat D_i^\dag(\SpatialArgs)$~\footnote{For completeness, and given the importance of coherent states in atomic physics and quantum chemistry we note that the displacement operator can also be written as $\hat D_i(\alpha) = \exp\left(\boldsymbol{\alpha}\cdot\hat{\boldsymbol{a}}^\dagger - \boldsymbol{\alpha^*}\!\cdot\hat{\boldsymbol{a}} \right) $ with parity $\SpatialKernel(\alpha) = \exp\left(\ui\pi\hat{\boldsymbol{a}^\dagger}\!\cdot\hat{\boldsymbol{a}}\right)$}. 
The generalized spin parity is $\SpinParity=(\Bid+\sqrt{3}\sigma_{z})/2$ and is chosen over a parity operator with eigenvalues $\pm$1 so that it satisfies Stratonovich-Weyl conditions~\cite{Rundle2016}.
The displacement operator for spin is $\hat{U}(\theta,\phi,\Phi)=\exp{(\ui \hat{\sigma}_z\phi)}\exp{(\ui \hat{\sigma}_y\theta)}\exp{(\ui \hat{\sigma}_z\Phi)}$ so that $\SpinKernel_i(\theta_i,\phi_i) =  \hat{U}_i(\theta_i,\phi_i,\Phi_i) \SpinParity \hat{U}_i^\dag(\theta_i,\phi_i,\Phi_i)$ for Euler angles $\theta_i$, $\phi_i$ (note that the third angle $\Phi_i$ cancels and plays no part in the Wigner function).
Given our focus on atomic physics and chemistry applications rather than quantum information,  a different sign convention is used for $\hat{U}(\theta,\phi,\Phi)$ and $\SpinParity$ to that used in~\Refs{PhysRevLett.117.180401,Rundle2016,Rundle2019} so that the Wigner function for $\sigma_z = +1$, i.e. spin up, points up.
A full discussion of this approach can be found in~\Ref{Rundle2016}.

The Wigner function for a composite system is found by taking expectation values of the tensor product of the displaced parity for each of the constituent parts.
The examples shown in~\Fig{Fig:Ref} provide a visual index of some important spin Wigner functions that will be used to inform later discussions, where the total spin parity is $\bigotimes_i\SpinKernel_i(\theta_i,\phi_i)$ over the appropriate set of spins.

For an $N$-electron atom, ignoring the nucleus, with density matrix $\hat\rho$ the Wigner function will be:
\be
    W(\mathbf{q}_1,\mathbf{p}_1,\theta_1,\phi_1,\ldots) = \Trace{\hat\rho\, \hat \Pi(\mathbf{q}_1,\mathbf{p}_1,\theta_1,\phi_1,\ldots)}, 
\ee
where
\be
    \hat \Pi(\mathbf{q}_1,\mathbf{p}_1,\theta_1,\phi_1,\ldots) = \bigotimes_{i=1}^{N} \ElectronParity_i(\SpatialArgs,\theta_i,\phi_i).
\ee
The generalized displaced parity for each electron has eight dimensions of which three are the spatial, $x_i$, $y_i$ and $z_i$, degrees of freedom, three are the concomitant momentum degrees of freedom and two are the spin degrees of freedom, $\theta_i$ and $\phi_i$. 
The Wigner function is therefore an $8N$-dimensional function --- distilling from this function meaningful visualizations of atomic states will be the subject of the next section.

How we choose to visualize the Wigner function depends very much on the application at hand. 
If, for example, the system is an electron in a periodic lattice, where momentum states are well defined, we might start by integrating out position degrees of freedom. 
This would yield a function that combines the probability density in the momentum representation with the spin Wigner function. 
If instead the system is an electron exposed to a potential that is periodic in one dimension and quadratic in perpendicular directions (such as a quantum wire or ion trap) it seems appropriate to integrate out the position degrees of freedom for the periodic component, and the momentum degrees of freedom for the other components. 
This would yield a function that combines the probability density function in the momentum representation for the periodic dimension, the position representation of the probability density and the spin Wigner function.

It is possible to extend our method to include the nucleus using a suitable spin-parity operator to represent the overall nuclear spin. 
The total atomic Wigner function is then obtained by taking expectation values of 
\be
    \hat \Pi_\mathrm{with\ nucleus}^\mathrm{He} =  \hat\Pi_\mathrm{nucleus}\otimes \ElectronParity_1 \otimes \ElectronParity_2,
\ee
which may be of interest for systems where the Jahn–Teller effect is important (see~\Refs{PhysRevLett.117.180401,Rundle2019} for details on how to construct $\hat\Pi_\mathrm{nucleus}$ for a given nuclear spin).
If more detail is required, displaced parity operators for protons and neutrons could be used so that  
\be
    \hat \Pi_\mathrm{total}^\mathrm{He} =  \hat \Pi_1^\mathrm{p^+} \otimes \hat \Pi_2^\mathrm{p^+}\otimes \hat\Pi_1^\mathrm{n}\otimes \hat\Pi_2^\mathrm{n}\otimes \ElectronParity_1 \otimes \ElectronParity_2.
\ee
If still more detail is required, it may even be possible to write the phase space representation for each nucleon's constituent parts (see~\Refs{PhysRevLett.117.180401,Rundle2019} for details on how to construct generalized displaced parity operators such as those needed for other spins and colour).

In a similar way, to describe an atom interacting with a field, or indeed molecules, the total parity is the tensor product of the parities of all the system's constituent parts.
This leads to a Wigner phase-space representation of the total quantum state.
%
\section{Results}
In this section we obtain a Wigner function visualization for a range of atomic states. 
At this stage, in order to simplify calculations, we use a model atom representation which replaces the Coulomb confining potential with that of a three-dimensional harmonic oscillator (as in~\Ref{cohen1977quantum}), and is similar in form to the Hooke and Moshinsky atoms in the non-interacting electron model~\cite{Coe2008,Garcia1998,Laguna2012,Osenda2007,Pipek2009,Yanez2009}. 
This approximation does not alter the angular distributions of the eigenstates and provides an adequate first approximation to the radial dependence of real hydrogenic systems which is sufficient for our present purposes. 
It has the additional advantage of allowing the calculation of momentum-only representations, such as are required for the visualization of Compton scattering profiles, for example~\Refs{Cooper1985,Singru1989}. 

 The states of hydrogen, helium and lithium referred to below are obtained within this approximation however, for simplicity, such states are referred to by their corresponding atomic name.
%
\subsection{Hydrogen}
Even though hydrogen is a one-electron system, the Wigner function is eight dimensional (with three spatial $\mathbf{q}$, three momentum $\mathbf{p}$, and two spin degrees of freedom).  
To produce from this a representation of hydrogen as similar as possible to existing images we integrate out the momentum degrees of freedom. 
\bel{hWFull}
    W^{\mathrm{H}}(\mathbf{q},\theta,\phi) := \int \ud^3 \mathbf{p}\, W^{\mathrm{H}}(\mathbf{q},\mathbf{p},\theta,\phi).
\ee
This results in a reduced Wigner function of only three spatial and two spin degrees of freedom. 
We adopt the notation throughout this work that the degrees of freedom not in the argument list have been integrated out resulting in a reduced Wigner function.
We now consider a visualization strategy that seeks to display as much of this information as is possible, whilst being constrained by our requirement to make this as familiar as possible. 

%
\begin{figure*}[!ht]
    \begin{tikzpicture}
        \node at (0,0){
        	\includegraphics[width = 1.0\linewidth, trim = {0cm, 0cm, 0cm, 0cm}, clip = true]{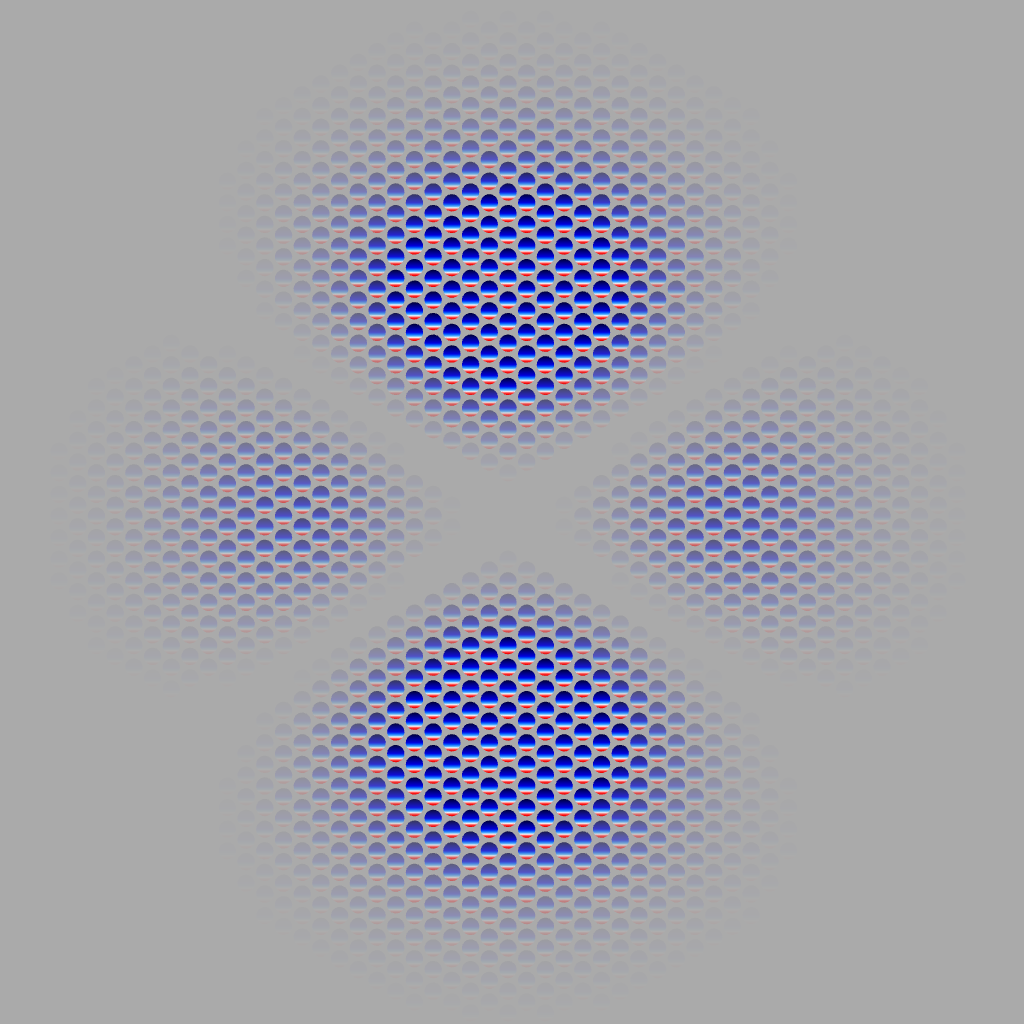}
         };
    
        \node at (6.35,-6.5) {
        	\includegraphics[width = 0.26\linewidth]{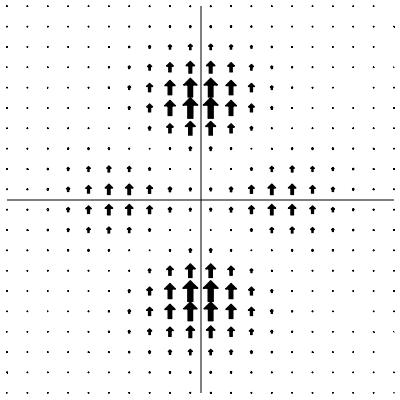}
         };
 
    \end{tikzpicture}
    \vspace{-20pt}\caption{
    This figure displays the spin up 3$d_{z^2}$ orbital for the three-dimensional harmonic oscillator. 
    The Wigner function for this orbital has 8 dimensions; the three spatial $x$, $y$, and $z$ degrees of freedom, the concomitant momentum degrees of freedom, and two spin degrees of freedom $\theta$ and $\phi$. 
    To obtain the familiar orbital structure, all momentum and spin degrees of freedom are integrated out to yield the probability density function in terms of position. 
    These values are used to set the opacity ($\alpha$) of each sphere, neglecting all points where $\alpha < 0.1$.  
    At each point, $\mathbf{q}$, in the $xz$-plane we plot the reduced Wigner function, $W^{\mathrm{H}}(\mathbf{q},\theta,\phi)$, on a sphere as in~\Fig{Fig:Ref} (see~\Eq{hWFull}).
    Each sphere can then be interpreted as an indication of the probability of finding an electron at $\mathbf{q}$ with a certain spin. 
    In this plot, which has rotational symmetry about the $z$ axis, the state of the system is of the same form as an $n=3$, $l=2$, $m=0$ $d$ orbital of hydrogen with spin pointing up (see~\FigSub{Fig:Ref}{a}).
    To aid interpretation, the inset shows an equivalent plot using arrows to represent the spin.
    }
\label{Fig:d}
\end{figure*}
%
\begin{figure*}[!ht]
    \begin{tikzpicture}

        \node at (0,0){
        	\includegraphics[width = 1.0\linewidth]{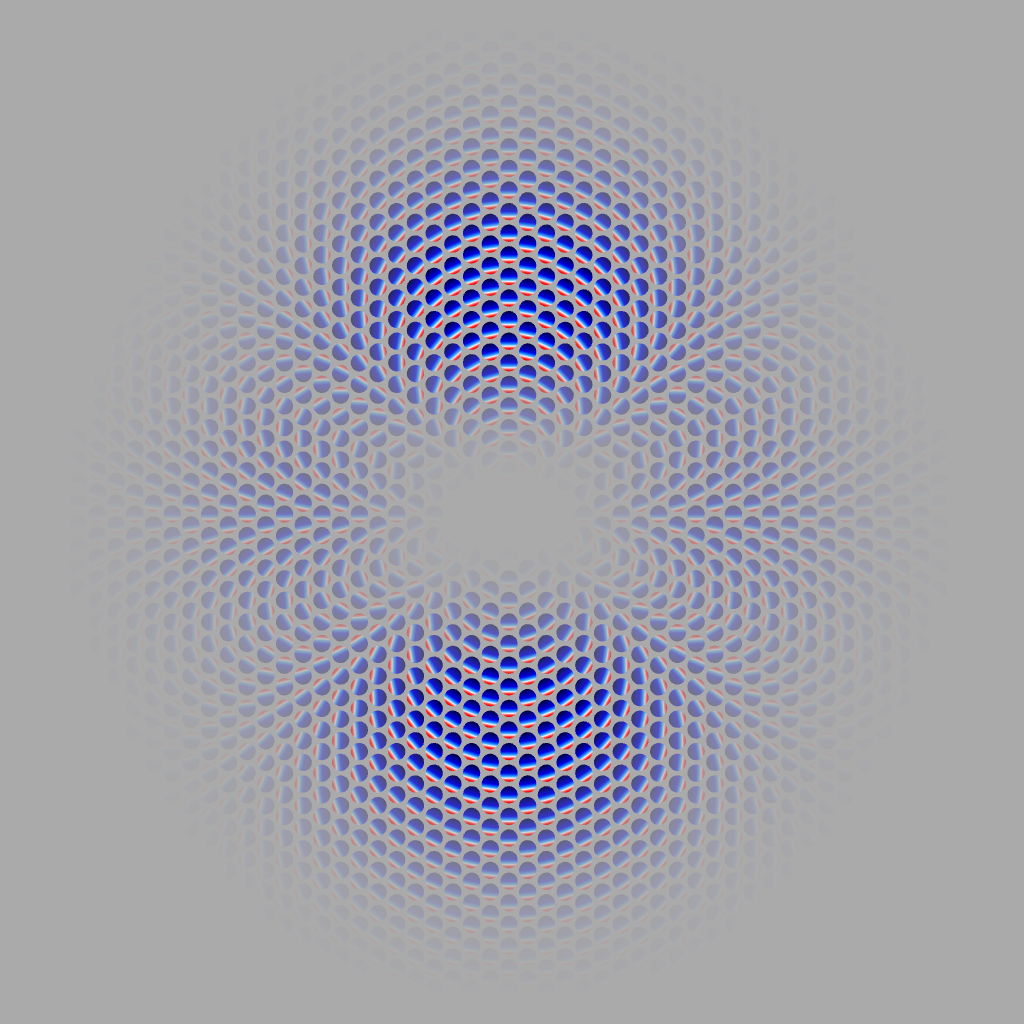}
         };
    
        \node at (6.3,-6.5){
        	\includegraphics[width = 0.26\linewidth]{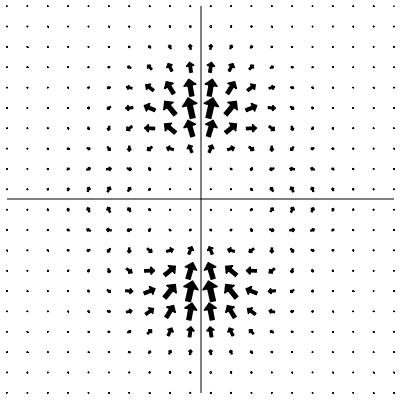}
            };
    
    \end{tikzpicture}
\caption{
    Due to relativistic effects in the Hamiltonian of real atomic hydrogen, states such as the one shown in~\Fig{Fig:d} are not stationary.
    One of the most important corrections arises due to a coupling between spin and orbital angular momentum degrees of freedom. 
    This affects every state, other than the $s$ orbitals, and the result is that the energy eigenstates have entangled spin and spatial degrees of freedom.
    Such entanglement cannot be made visible using conventional probability density plots. 
    This figure follows the same scheme as~\Fig{Fig:d} but for the $\ket{j=5/2, m=1/2}$ orbital; it is clear that there are correlations between the spin and spatial degrees of freedom.
    In this way we demonstrate how our method can visualize the entanglement of the electron's spin and orbital degrees of freedom, as the spin points in different directions at different positions.
    The inset shows an equivalent plot using arrows to represent the spin. 
}
\label{Fig:SO}
\end{figure*}

For the visualization we choose a set of points in space~\footnote{for simplicity, and with respect to symmetry, we have chosen a plane grid but this method can be extended to the three spatial dimensions}.
At each of these points a sphere is plotted with its opacity, $\alpha$, obtained from the value of 
\bel{modPsiH}
    |\psi^{\mathrm{H}}(\mathbf{q})|^2 =W^{\mathrm{H}}(\mathbf{q})= \frac{2}{\pi}\int_{0}^{\pi/2} \!\!\!\!\! \ud \theta\! \int_{0}^{\pi} \!\!\!\! \ud \phi \, \sin\!\left(2\theta\right)\, W^{\mathrm{H}}(\mathbf{q},\theta,\phi).
\ee
as $\alpha = W^{\mathrm{H}}(\mathbf{q}) / W^{\mathrm{H}}_{\mathrm{max}}(\mathbf{q}).$
This position marginal is simply the spatial probability density function. In order to more readily make comparison with standard orbital plots all spheres with an opacity less than $0.1$ have been omitted.
On the surface of the sphere at $\mathbf{q}$ is plotted the reduced Wigner function $W^{\mathrm{H}}(\mathbf{q},\theta,\phi)$.
This means that each sphere is an indication of the probability of finding an electron at that point in space with a certain spin. 

As a gentle introduction to our visualization scheme a simple state generated using the above scheme is plotted in~\Fig{Fig:d}.
The spatial dependence conforms to standard plots of $d_{z^2}$-orbitals of hydrogen.
Comparing each sphere with~\FigSub{Fig:Ref}{a}, the spin Wigner function at each point is consistent with the up state, $\ket{\uparrow}$.
From inspection we have been able to correctly infer that this is $\ket{d_{z^2},\uparrow}$
\footnote{
The basis states in this paper are represented as Fock states, where $\ket{n_x,n_y,n_z}$ are eigenstates of the three dimensional harmonic oscillator and $n_i$ indicates the number of photons in the $x$, $y$ or $z$ component, see ref.~\cite{cohen1977quantum} for details).
The forms of the relevant states are $\ket{d_{xz}}=\ket{101}$, $\ket{d_{yz}} = \ket{011}$, and
    $\ket{d_{z^2},\uparrow} = \sqrt{1/6}\left(2\ket{002}-\ket{200}-\ket{020}\right)\Up$
}.

Figure~\ref{Fig:SO} shows a less trivial state. 
It is interesting to explore what can be deduced from only this figure and~\Fig{Fig:Ref}.
The first observation is that the spheres are identical to that in~\FigSub{Fig:Ref}{a} but pointing in different directions. 
The more opaque spheres are predominantly pointing in one direction suggesting there is a corresponding overall magnetic moment.
Secondly, the direction of the spin varies as a function of position - this is an indication of correlation (entanglement) of the electron's spin and spatial degrees of freedom.
Neither of these two pieces of information are obtainable from conventional plots of atomic orbitals. 

In real atomic hydrogen the total energy is more than the sum of kinetic and Coulomb potential energies. 
There are a number of relativistic effects that need to be taken into account in order to get an accurate model that, for example, correctly predicts the energy level structure and thus the absorption/emission spectra of hydrogen.
One of the most important of these relativistic effects is the spin-orbit coupling term (proportional to $\hat{\mathbf{L}} \cdot \hat{\mathbf{S}}$).
It is not surprising therefore to find that the state represented in~\Fig{Fig:SO} is one such state. Specifically, 
\be
    \ket{j=\frac{5}{2},m=\frac{1}{2}} = \sqrt{\frac{3}{5}}\ket{d_{z^2}}\Up + \sqrt{\frac{1}{5}}(\ket{d_{xz}}+\ui\ket{d_{yz}})\Down,
\ee
which, as we deduced in our above discussion of~\Fig{Fig:SO}, has a non-zero magnetization ($1/2$), strongly entangles spin and spatial degrees of freedom and has an entropy of entanglement of 0.971 bits.
We note that the eigenstates \ket{j,m} are labelled by $j$ the quantum number associated with $\hat{J}^2=(\hat{\mathbf{L}}+\hat{\mathbf{S}})^2$ and $m$ the eigenvalue of $\hat{J}_z=\hat{L}_z+\hat{S}_z$ for orbital and spin angular momenta $\hat{\mathbf{L}}$ and $\hat{\mathbf{S}}$ respectively.  

%
\subsection{Helium}
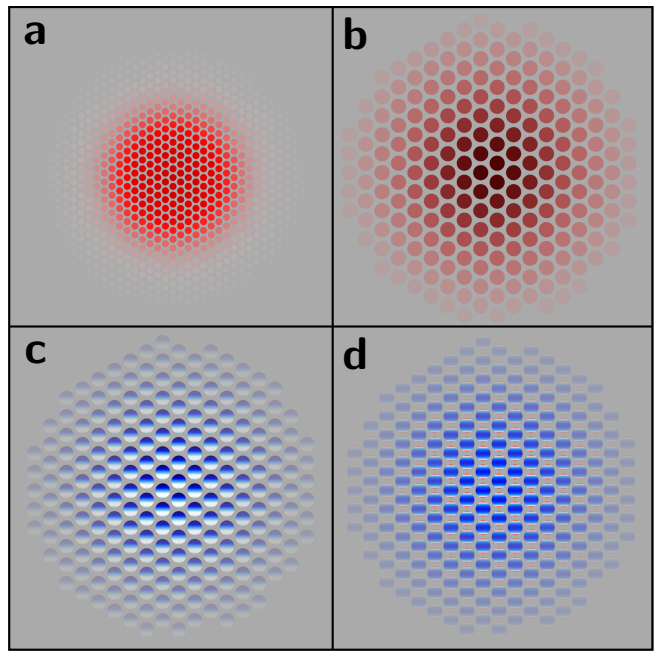
\begin{figure}[!h]
    \begin{tikzpicture}
        
        \node at (0.0,0.0){
	        \includegraphics[width = 0.5\linewidth, 
	            trim = {4cm, 4cm, 4cm, 4cm}, 
	            clip = true]
	            {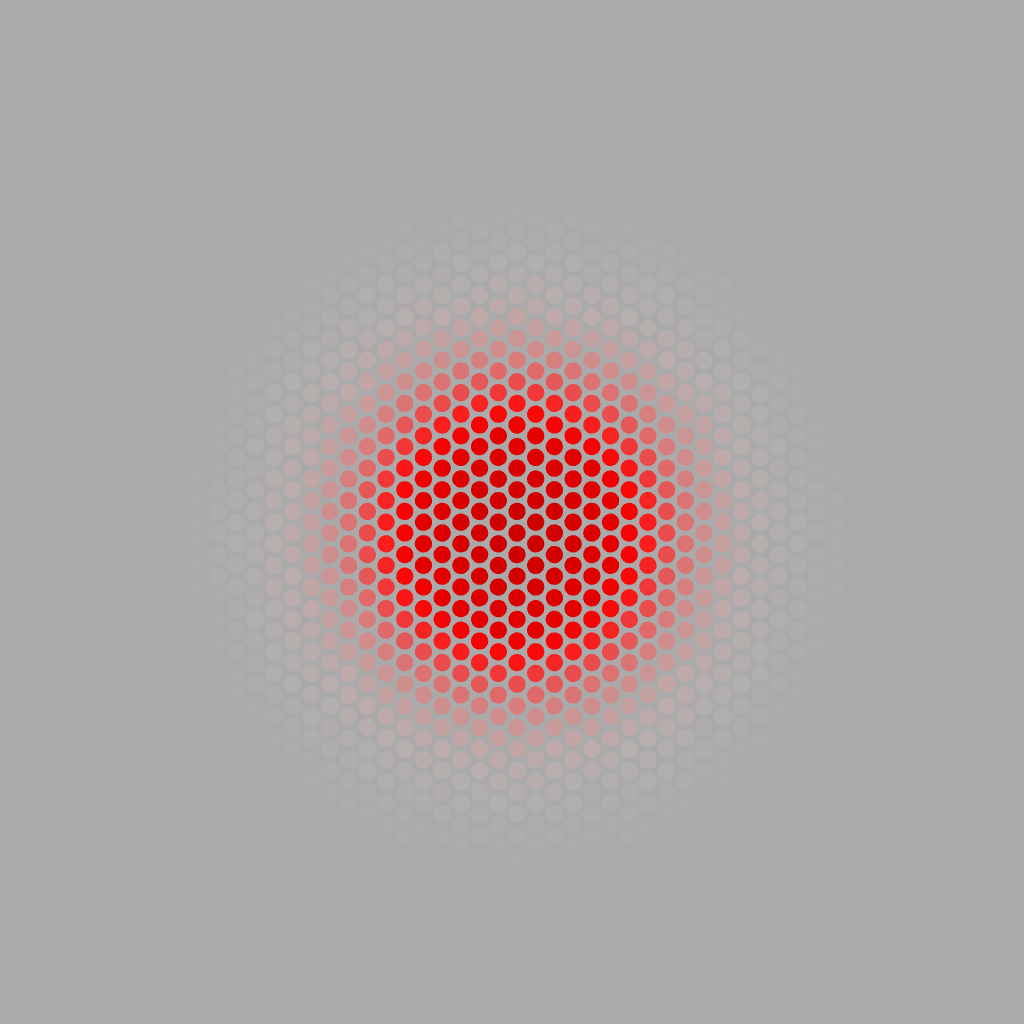}
        };
        
        \node at (4.25,0.0){
	        \includegraphics[width = 0.5\linewidth]
	            {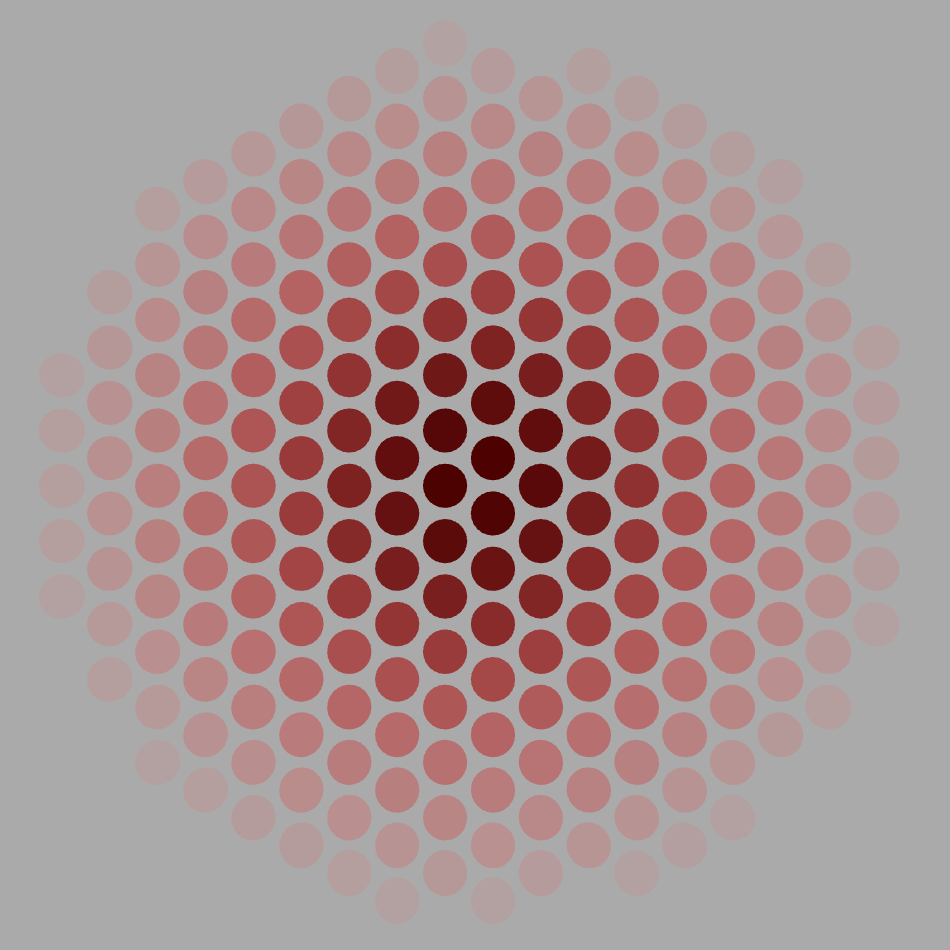}
        };
        
        \node at (0.0,-4.25){
	        \includegraphics[width = 0.5\linewidth]
	            {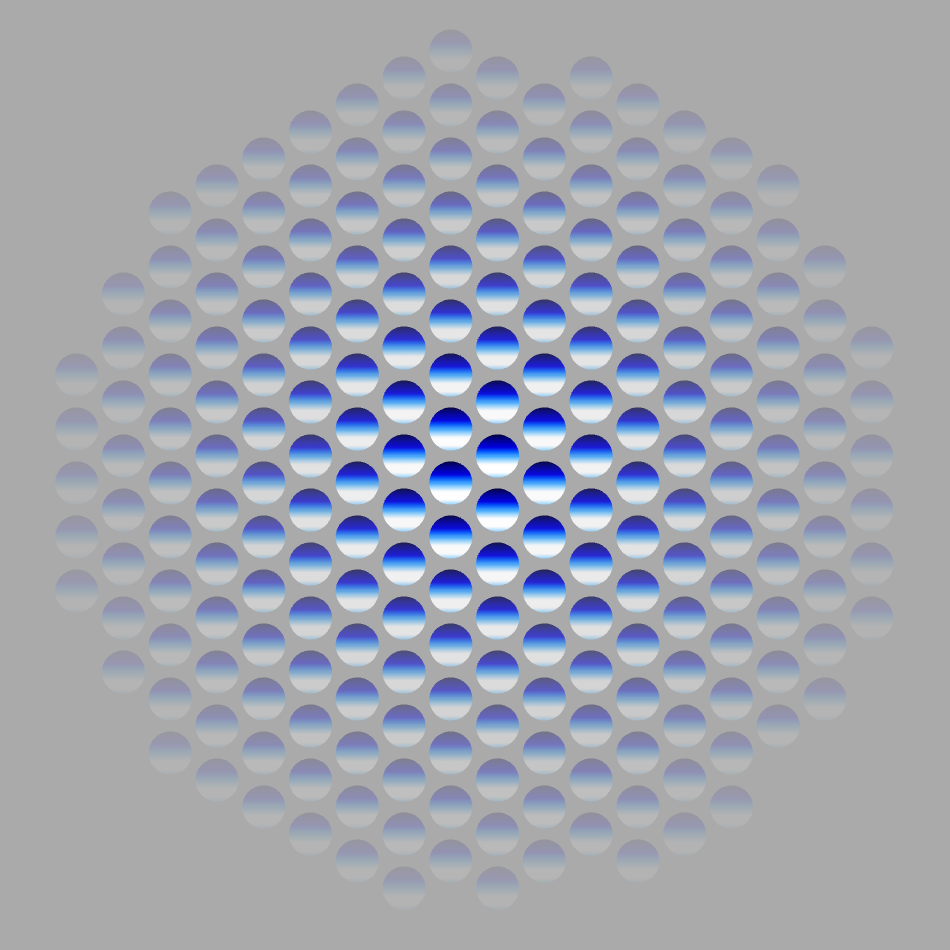}
        };
        
        \node at (4.25,-4.25){
	        \includegraphics[width = 0.5\linewidth]
	            {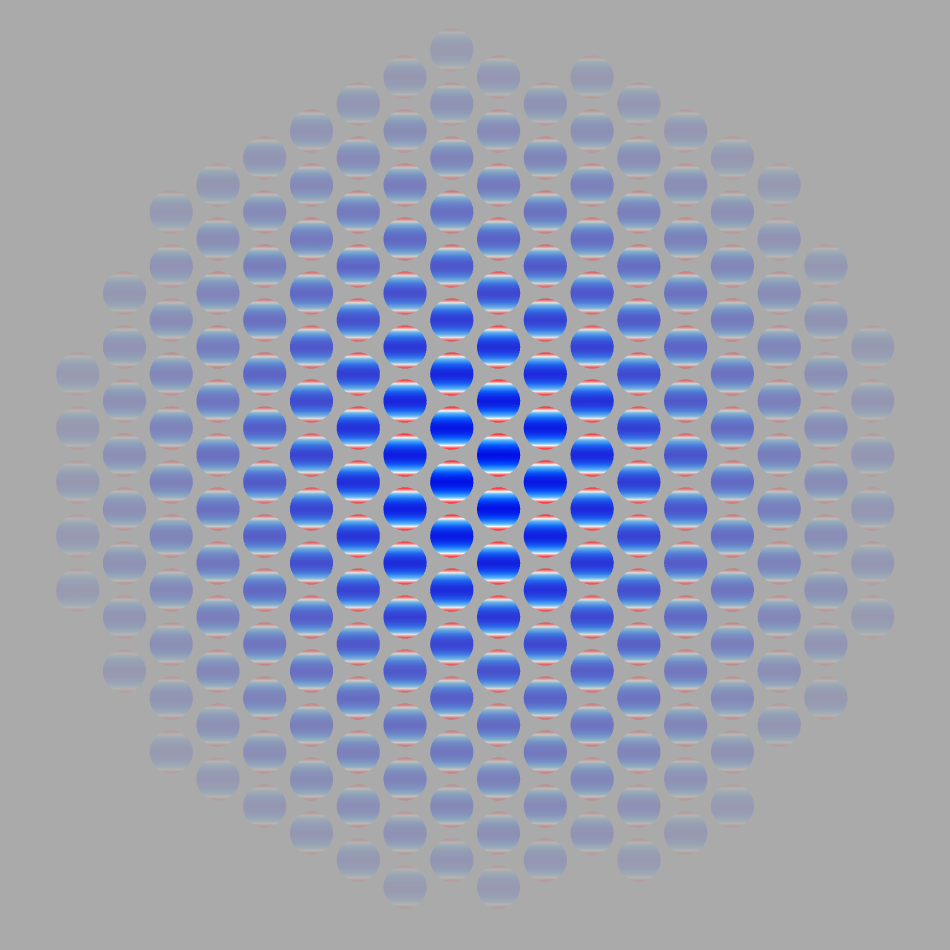}
        };

        \node at (-1.80,  1.80){\lettering{a}};
        \node at ( 2.45,  1.80){\lettering{b}};
        \node at (-1.80, -2.45){\lettering{c}};
        \node at ( 2.45, -2.45){\lettering{d}};

        \draw [thick] 
            (-2.15, 2.15) rectangle 
            (6.4,-6.4);
        \draw [thick] 
            (-2.15,-2.1) -- 
            (6.4,-2.1);
        \draw [thick] 
            (2.15, 2.15) -- 
            (2.15,-6.4);
    \end{tikzpicture}
\caption{
    This figure shows the equal-angle slice, $\theta_1=\theta_2=\theta$ and $\phi_1=\phi_2=\phi$, of the Wigner function for the following states of helium: \Sub{a}  ground state; \Sub{b}  first excited singlet; \Sub{c} first triplet state with magnetization quantum number $m=1$ (note for $m=-1$ each sphere would be the antipodal version of the ones shown here); \Sub{d} first triplet state with magnetization quantum number $m=0$.
    Comparing each figure with~\Fig{Fig:Ref} we see that \Sub{a} and~\Sub{b} correspond to the entangled state~\FigSub{Fig:Ref}{e}, and~\Sub{d} with the entangled state~\FigSub{Fig:Ref}{f}. 
    \Sub{c} corresponds to the non-entangled state in~\FigSub{Fig:Ref}{c}.
    In this way we demonstrate how our method not only clearly visualizes spin-orbit entanglement (as in~\Fig{Fig:SO}) but also spin-spin entanglement.
}
\label{Fig:Helium}
\end{figure}

We now begin to consider the case of multi-electron atoms. 
Helium's Wigner function is 16 dimensional having three spatial, three momentum and two spin degrees of freedom for each electron. 
To obtain the graphical representation of helium we use a similar scheme to the one used for hydrogen, also taking account of the Wigner function's increased dimensionality. 
Once more a reduced Wigner function is calculated $W^{\mathrm{He}}(\mathbf{q}_1,\theta_1,\phi_1,\theta_2,\phi_2)$, integrating out both electrons' momenta and one of the electron's spatial degrees of freedom (indistinguishability of electrons means that it will not matter which one is chosen).
Here the function $W^{\mathrm{He}}(\mathbf{q}_1) =|\psi^{\mathrm{He}}(\mathbf{q}_1)|^2$,  defined in the same manner as in~\Eq{modPsiH}, by integrating out all spin degrees of freedom, is again used to set the intensity.
In plotting multi-electron systems, we choose the equal angle slice of the Wigner function for the spin degrees of freedom, where $\theta_1 = \theta_2$ and $\phi_1 = \phi_2$.
The equal angle slice is a natural choice, as we want the Wigner function to remain the same upon permutation of indices due to the indistinguishability of electrons.
This slice is then plotted on the surface of each of the spheres in~\Fig{Fig:Helium} for helium.
%

In~\Fig{Fig:Helium} we have plotted the ground state,~\FigSub{Fig:Helium}{a}, the first excited singlet state,~\FigSub{Fig:Helium}{b}, and two of the triplet states,~\FigSub{Fig:Helium}{c} and~\Sub{d}, of helium.
In the ground state we see three key features:
(i) with reference to~\FigSub{Fig:Ref}{d}, each sphere is consistent with that of the two-spin singlet state (the antisymmetric superposition of spin up and spin down, and not \ket{\uparrow \downarrow} as in~\FigSub{Fig:Ref}{c}, often indicated in elementary treatments of the subject);
(ii) the intensity in this plot suggests the spatial component is the product of two $s$-orbitals and;
(iii) there is no dependence of spin on position, consistent with the spin and spatial degrees of freedom being separable.
These observations are consistent with the ground state of helium,  $\ket{1S(1)1S(2)}\left(\ket{\uparrow_1 \downarrow_2}-\ket{\downarrow_1 \uparrow_2 } \right)/\sqrt2$~\footnote{
This work has adopted the notation that the number before the $S$ (or $P$, $D$, \ldots) indicates the principal quantum number with the electron index in parentheses.}.
A comparison of the spins with~\Fig{Fig:Ref} for the remaining states demonstrates that both~\FigSub{Fig:Ref}{b} and~\Sub{d} are in an entangled spin state, whilst~\Sub{c} is not.

\subsection{Lithium}
%
\begin{figure*}[!t]
    \begin{tikzpicture}
        \definecolor{BackgroundColour}{rgb}{0.66667, 0.66667, 0.66667}
        \def\ledg{-9.85}
        \def\redg{7.9}
        \def\tedg{13.6}
        \def\bedg{-7.24}
        \def\yDivs{6.95}
        \def\diaglineYa{\tedg-\yDivs}
        \def\diaglineYb{\tedg-2*\yDivs}
        \fill [BackgroundColour] (\ledg,\diaglineYa) rectangle (\redg,\bedg);

        \node at (-5.3,\tedg-1.5*\yDivs){
        	\includegraphics[width = 0.45\linewidth, 
        	    trim = {4cm, 6cm, 4cm, 6cm}, 
        	    clip = true]
        	    {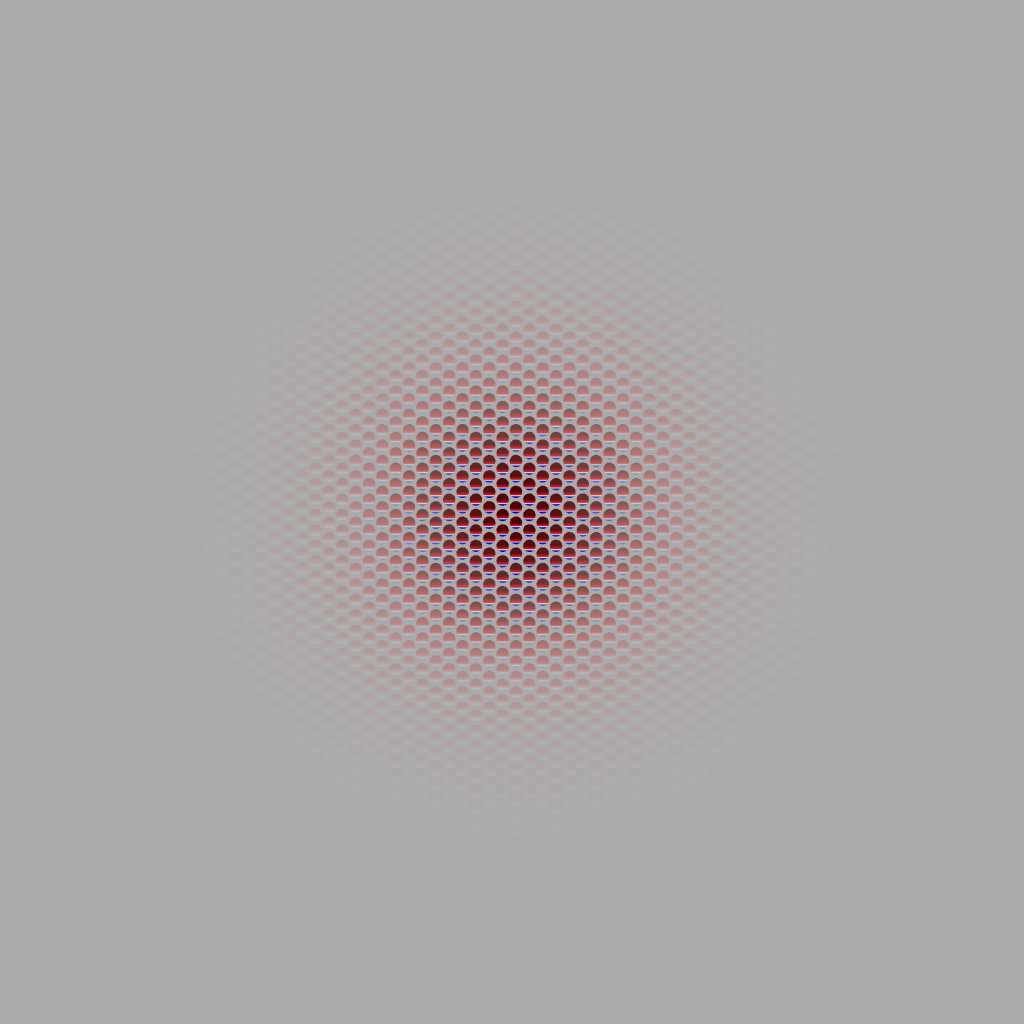}
         };
        \node at (3.43,\tedg-1.5*\yDivs){
        	\includegraphics[width = 0.45\linewidth, 
        	    trim = {4cm, 6cm, 4cm, 6cm}, 
        	    clip = true]
        	    {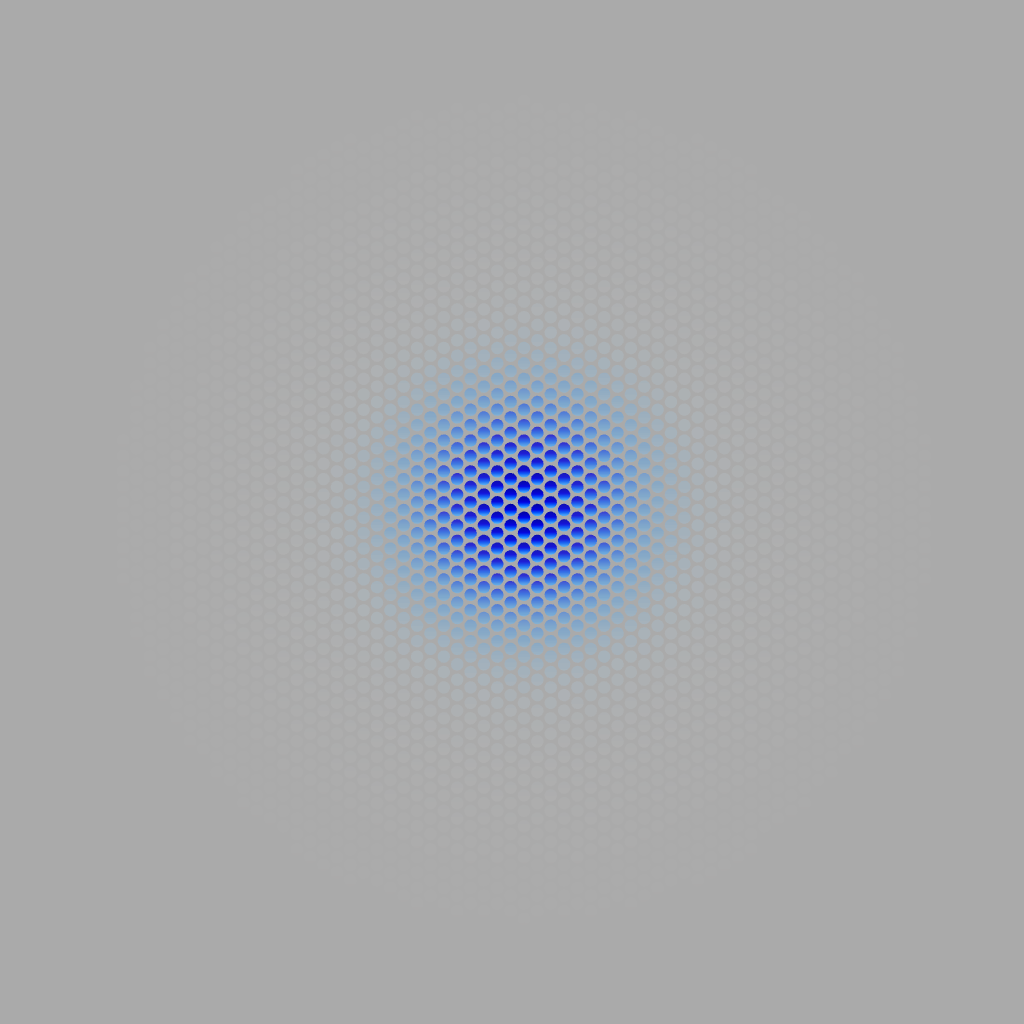}
         };
    
        \node at (-5.4,\tedg-2.5*\yDivs){
        	\includegraphics[width = 0.495\linewidth, 
        	    trim = {4cm, 7.08cm, 4cm, 7.08cm}, 
        	    clip = true]
        	    {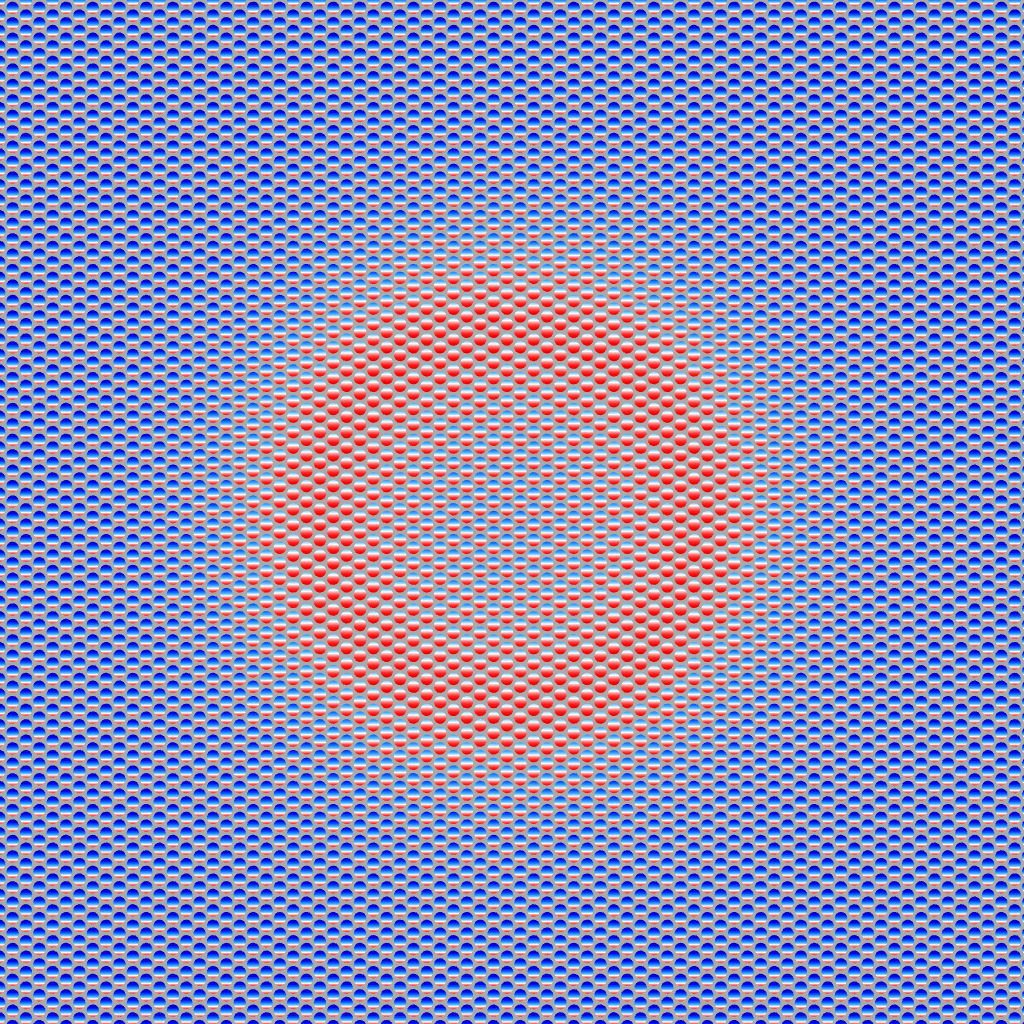}
         };
        \node at (3.43,\tedg-2.5*\yDivs){
        	\includegraphics[width = 0.495\linewidth, 
        	    trim = {4cm, 7.08cm, 4cm, 7.08cm}, 
        	    clip = true]
        	    {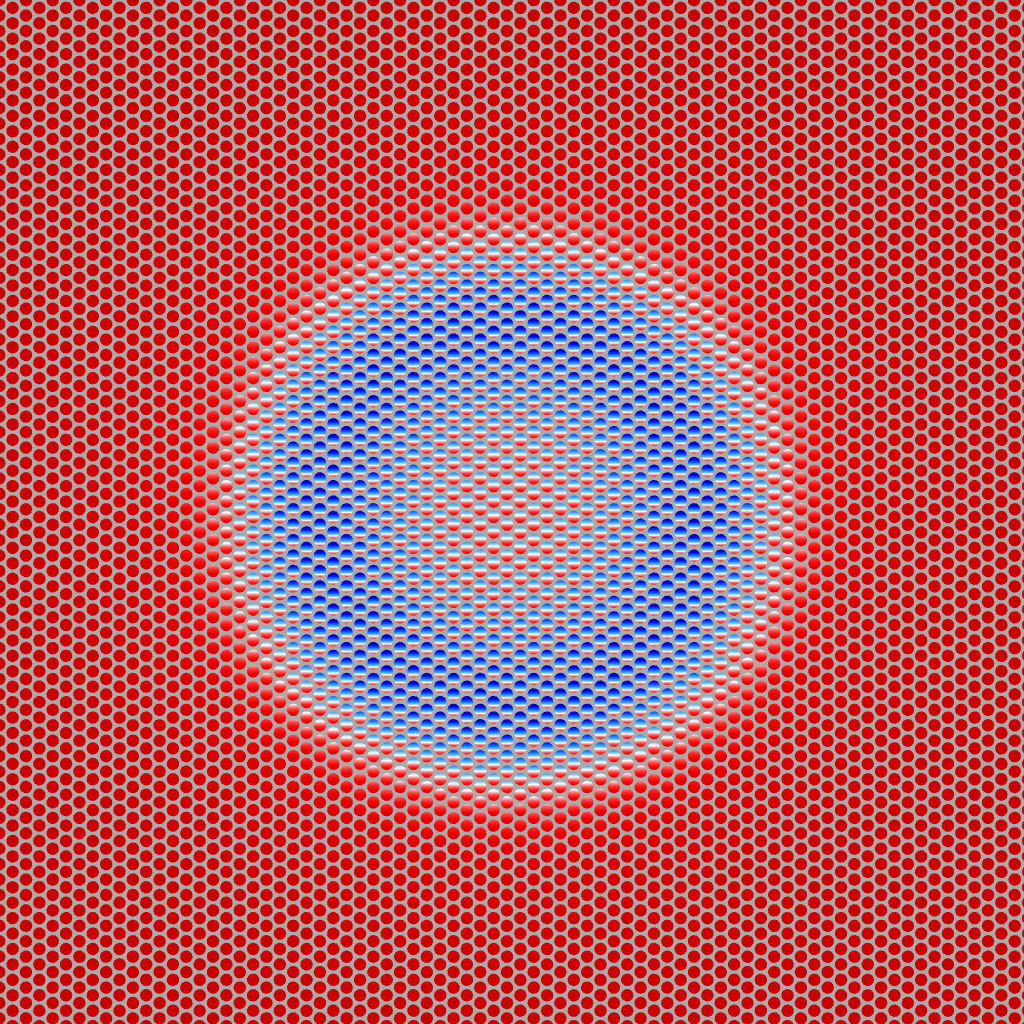}
        };
        \def\lineDispWhite{0.03}

        \def\HHeX{\ledg/2+\redg/2}
	
        \draw [thick,white] (\ledg-\lineDispWhite,\diaglineYa+\lineDispWhite) rectangle (\redg+\lineDispWhite,\bedg-\lineDispWhite);
        \draw [thick] (\ledg,\tedg-\yDivs) rectangle (\redg,\bedg);
	    \draw [thick] (\ledg,\diaglineYb) -- (\redg,\diaglineYb);
        \draw [thick] (\HHeX,\diaglineYa) -- (\HHeX,\bedg);
        
        
        \draw [white,very thick] (3.43,\tedg-2.5*\yDivs) circle (2.15);

    	\def\letterXDisp{0.6}
        \def\letterYDisp{0.6}

        \node at (\ledg+\letterXDisp, \diaglineYa - \letterYDisp){\lettering{a}};
        \node at (\HHeX+\letterXDisp, \diaglineYa - \letterYDisp){\lettering{b}};
        \node at (\ledg+\letterXDisp, \diaglineYb - \letterYDisp){\letteringW{c}};
        \node at (\HHeX+\letterXDisp, \diaglineYb - \letterYDisp){\letteringW{d}};
        \node at (6.0, \tedg - 2.5*\yDivs){\letteringW{Y}};
        \node at (7.0, \tedg - 2.5*\yDivs - 2.5){\letteringW{X}};
    
    \end{tikzpicture}
\caption{
    Showcasing the power of the Wigner function we demonstrate how to reconstruct all the important aspects of the Slater determinant for lithium by inspection of different slices (these figures are on a different scale to others to accommodate the $2S$ orbital).
    We follow the same scheme as in~\Fig{Fig:Helium}, on the equal angle slice where appropriate.
    In~\Sub{a} is the reduced Wigner function $W^{\mathrm{Li}}\!\left(\mathbf{q}_1,\theta_1,\phi_1,\theta_2,\phi_2,\theta_3,\phi_3\right)$ which at the origin is similar to that displayed in~\FigSub{Fig:Ref}{h}.
    Importantly, this shows that the spin entanglement structure in~\FigSub{Fig:Ref}{h} is part of the state. 
    In~\Sub{b} we extract the electron spin density, plotting the reduced Wigner function $W^{\mathrm{Li}}\!\left(\mathbf{q}_1,\theta_1,\phi_1\right)$. 
    This means that lithium must have an overall magnetic moment and, by comparison with~\FigSub{Fig:Ref}{a}, we see this manifested as the preponderance of blue in the positive $z$-direction.
    In~\Sub{c} and~\Sub{d} we have removed the link between transparency and amplitude of the position marginal to explore some of the more complex aspects of the quantum correlations.
    \Sub{c} is the reduced Wigner function $W^{\mathrm{Li}}\!\left(\mathbf{q}_1,\theta_1,\phi_1,\theta_2,\phi_2\right)$. 
    Note that integrating out $\theta_2$ and $\phi_2$ instead yields the same result, as the only spatial component is $\mathbf{q}_1$.
    \Sub{d} is the reduced Wigner function $W^{\mathrm{Li}}\!\left(\mathbf{q}_1,\theta_2,\phi_2,\theta_3,\phi_3\right)$. 
    At point \textsf{X} we find the singlet state $\ket{\uparrow_2\downarrow_3} - \ket{\downarrow_2\uparrow_3}$ which when combined with the state at a similar point in~\Sub{c} leaves the electron associated with $\mathbf{q}_1$ as spin up; this is consistent with $\ket{\uparrow_1}(\ket{\uparrow_2\downarrow_3} - \ket{\downarrow_2\uparrow_3})$.
    In the node of the $2S$ orbital (indicated by the ring \textsf{Y}) the spins form a mixed state, as we have integrated out entangled degrees of freedom, and in~\Sub{c} they form a singlet.
    This means that when $\mathbf{q}_1$ must be in the $1S$ orbital it is in a singlet state.
    Putting this together we deduce a state consistent with $\ket{2S(1),1S(2),1S(3)} \left(\ket{\uparrow_2\downarrow_3}-\ket{\downarrow_2\uparrow_3}\right)\ket{\uparrow_1}$.
    Coupled with the fact that the pictures must be invariant under cyclic permutation of electron indices (Pauli's exclusion principle) we infer that the state is $\ket{\psi^\mathrm{Li}} = \frac{1}{\sqrt{6}} [ \ket{1S(1),1S(2),2S(3)} \left(\ket{\uparrow_1\downarrow_2}-\ket{\downarrow_1\uparrow_2}\right)\ket{\uparrow_3} + \ket{1S(1),2S(2),1S(3)} \left(\ket{\downarrow_1\uparrow_3}-\ket{\uparrow_1\downarrow_3}\right)\ket{\uparrow_2} + \ket{2S(1),1S(2),1S(3)} \left(\ket{\uparrow_2\downarrow_3}-\ket{\downarrow_2\uparrow_3}\right)\ket{\uparrow_1} ].$
}
\label{Fig:Li}
\end{figure*}
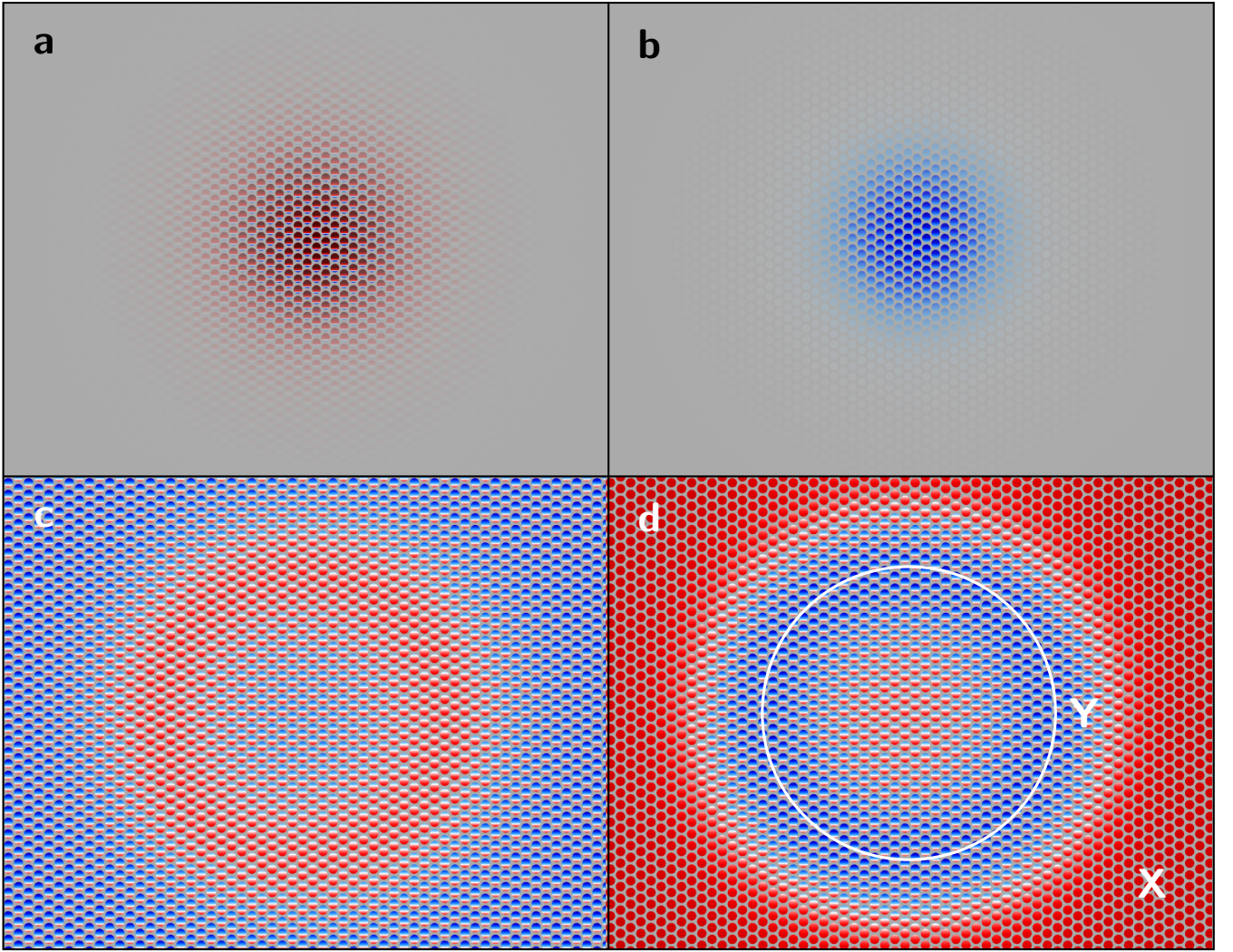
%
As with helium, lithium is often introduced along the following simplified lines:  two electrons are added to the $1S$ orbital with opposite spin, as dictated by the Pauli exclusion principle.
It also states that the third electron cannot be in the $1S$ orbital as it is now fully occupied. 
This electron must therefore go into the $2S$ orbital with spin \Up for example. 
The actual configuration of electrons in lithium is not this simple. 

The state of multi-fermionic systems can be found using the Slater determinant which ensures that Pauli's exclusion principle is properly satisfied and for lithium is 
\begin{equation}
    \ket{\psi^\mathrm{Li}} = \frac{1}{\sqrt{3!}}
    \begin{vmatrix}
        \ket{1S(1)}\ket{\uparrow_1} & \ket{1S(1)}\ket{\downarrow_1} & \ket{2S(1)}\ket{\uparrow_1}    \\ 
        \ket{1S(2)}\ket{\uparrow_2} & \ket{1S(2)}\ket{\downarrow_2} & \ket{2S(2)}\ket{\uparrow_2}    \\ 
        \ket{1S(3)}\ket{\uparrow_3} & \ket{1S(3)}\ket{\downarrow_3} & \ket{2S(3)}\ket{\uparrow_3}  
    \end{vmatrix},
\end{equation}
yielding,
\begin{align}
    \label{slater}
    \begin{split}
    	\ket{\psi^\mathrm{Li}} = \frac{1}{\sqrt{6}} &[ \ket{1S(1)1S(2)2S(3)} \left(\ket{\uparrow_1\downarrow_2} - \ket{\downarrow_1\uparrow_2}\right)\ket{\uparrow_3}  \\
            &+ \ket{1S(1)2S(2)1S(3)} \left(\ket{\downarrow_1\uparrow_3}-\ket{\uparrow_1\downarrow_3}\right)\ket{\uparrow_2}   \\
            &+ \ket{2S(1)1S(2)1S(3)} \left(\ket{\uparrow_2\downarrow_3}-\ket{\downarrow_2\uparrow_3}\right)\ket{\uparrow_1} ]
        \end{split}
\end{align}
or
\begin{align}
    \begin{split}
    	= \frac{1}{\sqrt{6}} &[ \ket{\uparrow_1\uparrow_2\downarrow_3} \left(\ket{2S(1)1S(2)}-\ket{1S(1)2S(2)}\right)\ket{1S(3)}    \\
            &+ \ket{\uparrow_1\downarrow_2\uparrow_3} \left(\ket{1S(1)2S(3)}-\ket{2S(1)1S(3)}\right)\ket{1S(2)} \\
            &+  \ket{\downarrow_1\uparrow_2\uparrow_3} \left(\ket{2S(2)1S(3)}-\ket{1S(2)2S(3)}\right)\ket{1S(1)}].
    \end{split}
\end{align}

The ground state of lithium is a superposition of all the possible Slater determinants but here we shall only consider this one.
From~\Eq{slater}, it can be seen that there is bipartite entanglement between each spin degree of freedom.
There is also a non-trivial level of spin-spatial entanglement combining these bipartite entangled spin states. 
Entanglement such as this could be an important factor in determining physical and chemical properties~\cite{Boguslawski2015,Dehesa2012, Esquivel2011,Osenda2007, Tichy2011}.
Therefore, being able to get a grasp of such phenomena without necessarily analyzing the full mathematics would be of tremendous value.
We now explore an example of how our visualization strategy can be utilized in achieving such an ambition.

Lithium has a 24-dimensional Wigner function (the usual eight dimensions for each electron). 
Due to the added complexity of lithium, it is now necessary to look at different slices of the Wigner function. 
As before all momentum degrees of freedom have been integrated out, however spin degrees of freedom have also been integrated out, appropriate to each figure.
For those slices with multiple electron spin degrees of freedom remaining, the equal angle slice is used.
We show a selection of different slices in~\Fig{Fig:Li}.
Although we have restricted this discussion to the four slices presented, other slices could be chosen to explore different features of the state. 

In~\FigSub{Fig:Li}{a}, the spatial degrees of freedom $\mathbf{q}_2,$ and $\mathbf{q}_3$ have been integrated out.
This leaves the reduced Wigner function $W^{\mathrm{Li}}\!\left(\mathbf{q}_1,\theta_1,\phi_1,\theta_2,\phi_2,\theta_3,\phi_3\right)$.
The function behaviour at the origin of~\FigSub{Fig:Li}{a} is similar to that displayed in~\FigSub{Fig:Ref}{h}. 
It is important to note that the state differs from~\FigSub{Fig:Ref}{h} because what is shown is not itself pure.
The reason for it being mixed is that this is a single slice of the full Wigner function with entangled degrees of freedom integrated out.
Points far from the origin tend towards the pure variation of~\FigSub{Fig:Ref}{h}, where an electron is in the up state and likely to be found in the $2S$ orbital.
This slice is consistent with the description of lithium as a singlet state in the $1S$ orbital coupled with a spin up in the $2S$ orbital.

Figure~\ref{Fig:Li}~\Sub{b} is a plot of the reduced Wigner function $W^{\mathrm{Li}}\!\left(\mathbf{q}_1,\theta_1,\phi_1\right)$.
This slice gives us insight into the electron spin density, revealing the magnetization of lithium.
Lithium has an overall magnetic moment which is manifested as the preponderance of blue in the up direction (compare with~\FigSub{Fig:Ref}{a}).
There is no negativity in this plot as a sufficient amount of entanglement information has been integrated out to produce a Wigner function of a mixed state. 

Figures~\ref{Fig:Li}~\Sub{c} and~\Sub{d} explore some of the more complex aspects of the quantum correlations within lithium, that combine both spin-spin and spin-orbit entanglement.
To study these entanglement effects in more detail, we have removed the link between transparency and amplitude of the position marginal.

Figure~\ref{Fig:Li}~\Sub{c} is the equal-angle slice of the reduced Wigner function $W^{\mathrm{Li}}\!\left(\mathbf{q}_1,\theta_1,\phi_1,\theta_2,\phi_2\right)$. 
We note that integrating out $\theta_2$ and $\phi_2$ instead of $\theta_3$ and $\phi_3$ yields the same result, as the only spatial component is $\mathbf{q}_1$. 
The region dominated by red is the node of the $2S$ orbital and implies that if the electron associated with $\mathbf{q}_1$ is found here it is likely to be in a singlet state.
%
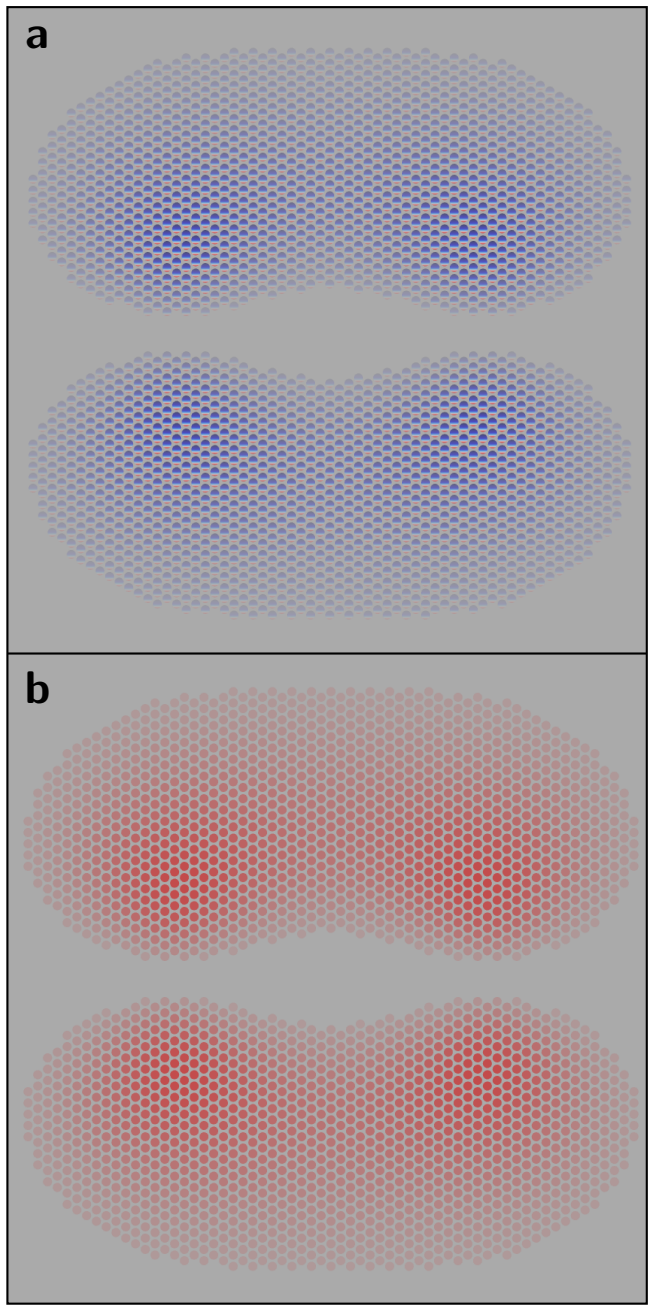
\begin{figure}[!t]
    \begin{tikzpicture}
    \definecolor{BackgroundColour}{rgb}{0.66667, 0.66667, 0.66667}
        \fill [BackgroundColour] (-4.3,4.3) rectangle (4.2,-12.95);
        
        \node at (0.0,0.0){
	        \includegraphics[width = 0.95\linewidth]
	            {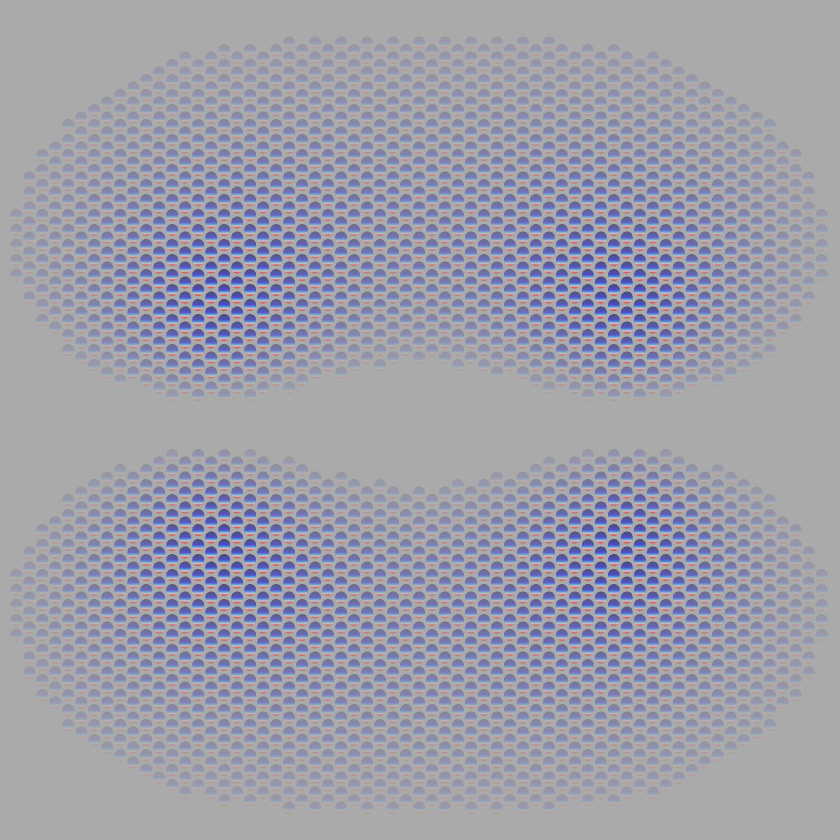}
        };

        \node at (0.0,-8.65){
	        \includegraphics[width = 0.97\linewidth]
	            {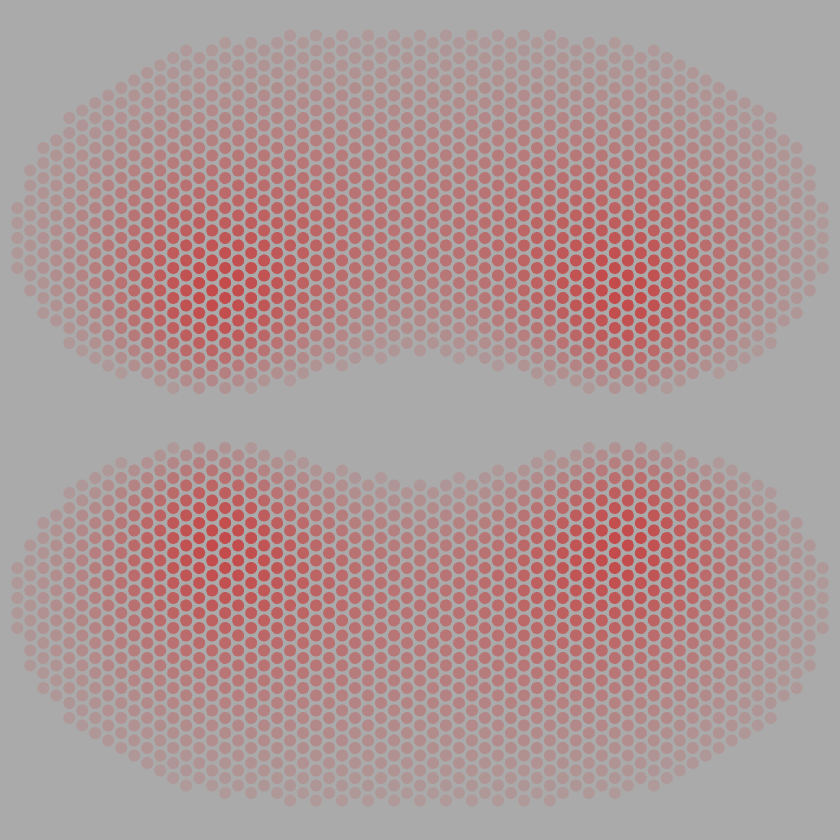}
        };

        \node at (-3.9,  3.9){\lettering{a}};
        \node at (-3.9,  -4.75){\lettering{b}};
    
        \draw [thick] 
            (-4.3, 4.3) rectangle 
            (4.2,-12.95);
        \draw [thick] 
            (-4.3,-4.3) -- 
            (4.2,-4.3);
    \end{tikzpicture}
\caption{
Simplified versions of single electron, \FigSub{Fig:Molecules}{a}, and double electron, \FigSub{Fig:Molecules}{b}, $\pi-$bonds in a $p$-bonded pseudo-molecule. 
Note that in the linear combination of atomic orbitals approximation the spatial components are identical, the states can only be visually distinguished through spin degrees of freedom - this difference is clearly seen in the Wigner functions displayed above.
States where this distinction is important will arise often in organic chemistry.
}
\label{Fig:Molecules}
\end{figure}

Figure~\ref{Fig:Li}~\Sub{d} is the equal-angle slice of the reduced Wigner function $W^{\mathrm{Li}}\!\left(\mathbf{q}_1,\theta_2,\phi_2,\theta_3,\phi_3\right)$.
Here we see that if the electron associated with $\mathbf{q}_1$ is far from the origin, the other two electrons are likely to form a singlet.
By forming a singlet the electrons have high probability of being in the same orbital, the $1S$ orbital.
Furthermore, where the $2S$ contribution is close to zero, there is little contribution from the singlet state indicated by the lack of negativity. 
Hence, the electrons associated with $\mathbf{q}_2$ and $\mathbf{q}_3$ are not likely to be in the same orbital at these points.

Putting all this together, and taking recognition of the permutations, we see from~\Fig{Fig:Li} that we can infer the Slater determinant, and get substantial insight into advanced aspects of the quantum nature of lithium. 
This analysis is performed purely on the basis of the supporting table of spin-Wigner function reference states,~\Fig{Fig:Ref}.

\section{Molecules}
The importance of including spin degrees of freedom in the visualization of atoms and molecules is clearly illustrated in~\Fig{Fig:Molecules} which shows simplified versions of single electron, \FigSub{Fig:Molecules}{a}, and double electron, \FigSub{Fig:Molecules}{b}, $\pi-$bonds. 
The spatial distributions of these two pseudo-molecules are identical in the linear combination of atomic orbitals approximation~\cite{haaland2008molecules}. 
However the spin provides a distinguishing feature in the visualization for each state. 
Such situations will naturally be important in organic chemistry.

We note that a full quantum mechanical calculation of real molecular bonds including terms from spin-spin, spin-orbit, electron-electron, nuclear interaction, other relativistic effects etc., will have a substantial effect on the forms of these Wigner functions. 
As such \FigSub{Fig:Molecules}{a} and~\Sub{b} provide only a first glimpse of the potential that Wigner functions have for understanding the role of spin and entanglement in chemical processes.
However such analysis is beyond the scope of this paper and will be considered in future work.

\section{Concluding Remarks}
In this work we have shown that is possible to visualize various forms of atomic entanglement in an accessible way. 
Specifically, we have considered spin-orbit coupling (in hydrogen), spin only entanglement (in helium), and more complex hybrid entanglement (in lithium).
Importantly, we have been able to infer each of the states from the visualization alone.
We believe that this visualization technique will be of great utility in communicating the more complex and subtle aspects of the quantum mechanics of atoms and molecules, not just within the professional scientific community but also beyond.
We note that the Wigner function is found by taking expectation values of displaced parity operators each of which commute with one another and are observables.
Should simultaneous measurement of these quantities be possible, then the direct measurement of the system's Wigner function could be considered a form of quantum state spectroscopy.

%
\acknowledgments
TT notes that this work was supported in part by JSPS KAKENHI (C) Grant Number JP17K05569. 
RPR is funded by the EPSRC [grant number EP/N509516/1].  
MJE and TT thank Gergely Juhasz  and Steve Christie for interesting and informative discussions.
All authors thank Pooja Goddard (n\'ee Panchmatia) for informative discussions.
%
\bibliographystyle{apsrev4-1}
\bibliography{refs}
\end{document}